\documentclass[preprint,aps,nofootinbib,showpacs]{revtex4}
\input{epsf}

\usepackage{amsmath}
\usepackage{ulem}

\newcommand{\as}{\alpha_s}
\newcommand{\bpi}{{\bar \pi}}
\newcommand{\pimn}{\pi_{\mu\nu}}

\begin{document}
\title{Shear viscosity and out of equilibrium dynamics}

\author{Andrej El$^1,$\footnote{el@th.physik.uni-frankfurt.de}, Azwinndini Muronga$^{2,}$\footnote{Azwinndini.Muronga@uct.ac.za}, Zhe Xu$^{1,}$\footnote{xu@th.physik.uni-frankfurt.de}, Carsten Greiner$^{1,}$\footnote{Carsten.Greiner@th.physik.uni-frankfurt.de}}

\affiliation{$^1$ Institut f\"{u}r Theoretische Physik, Goethe-Universit\"{a}t Frankfurt, Max-von-Laue Strasse 1, D-60438, Frankfurt am Main, Germany}
\affiliation{$^2$ Institute for Theoretical Physics and Astrophysics, Department of Physics, University of Cape Town,
Rondebosch 7701, South Africa; UCT-CERN Research Centre, Department of Physics, University of Cape Town, Rondebosch 7701, South Africa}

\begin{abstract}
Using  Grad's method, we calculate the entropy production and derive 
a formula for the second-order shear viscosity coefficient in a one-dimensionally expanding 
 particle system, which can also be considered out of chemical equilibrium. 
For a one-dimensional expansion of gluon matter with Bjorken boost 
invariance, the shear tensor and the shear viscosity to entropy density 
ratio $\eta/s$ are numerically calculated by an iterative and self-consistent
prescription within the second-order Israel-Stewart hydrodynamics and 
by a microscopic parton cascade transport theory. Compared with
$\eta/s$ obtained using the Navier-Stokes approximation, the present result 
is about $20\%$ larger at a QCD coupling $\alpha_s \sim 0.3$(with $\eta/s\approx 0.18$) and is 
a factor of $2-3$ larger at a small coupling $\alpha_s \sim 0.01$. We demonstrate an agreement
between the viscous hydrodynamic calculations and the microscopic transport results on $\eta/s$, 
except when employing a small $\alpha_s$. 
On the other hand, we demonstrate that 
for such small $\alpha_s$, the gluon system is far from kinetic and chemical 
equilibrium, which indicates the break down of second-order hydrodynamics because of the strong noneqilibrium evolution. 
In addition, for large 
$\alpha_s$ ($0.3-0.6$), the Israel-Stewart hydrodynamics formally breaks down at large momentum $p_T\gtrsim 3$ GeV but is still a reasonably good approximation.

\end{abstract}

\pacs{24.10.Lx, 24.10.Nz,12.38.Mh, 25.75.-q, 05.60.-k}

\maketitle

\section{Introduction}
Recent experimental measurements on the elliptic flow parameter $v_2$ at
the BNL Relativistic Heavy Ion Collider (RHIC) \cite{star,phen,phobos} show
a strong collectivity of the deconfined quark-gluon matter. The matter
produced was thus specified as a strongly coupled quark-gluon plasma
(sQGP) \cite{TDLee01,GM05,Shur05} or as a perfect 
fluid \cite{H01}. Further attempts to determine how imperfect 
the sQGP really is have drawn attention to transport coefficients like
viscosity \cite{Baym90,Hosoya85,Heiselberg94,AYM,XUPRL,HG06,Meyer07,KSS04,KNK09}
and to the derivation and solution of viscous 
hydrodynamics \cite{br07,Teaney03,SongHeinzCh06,hs08,Hama01,AM207,R97,DT08},
which is still a mathematical challenge.

Most current viscous hydrodynamic equations are based on second-order
Israel-Stewart kinetic theory \cite{IS}. They are solved numerically
using the given viscosity coefficients and initial conditions as well as
parton and hadron equation of state. In particular, the shear viscosity
to entropy density ratio $\eta/s$ is determined by comparing
the elliptic flow from the viscous hydrodynamical calculations 
 with the data at RHIC, as has been done recently in Refs. \cite{Romatschke08,RomPRL08}, 
where the value $\eta/s \approx 0.1$ was obtained.
On the other hand, even though the early partonic phase 
may be well described by ideal hydrodynamics ($\eta=0$), the hadronic afterburning \cite{Hirano06} has a larger dissipative effect, which may be enough 
to slow down the generation of the elliptic flow and bring its final value
into agreement with the data.

Dissipative phenomena can be alternatively described in transport calculations
solving Boltzmann equations of matter constituents \cite{Geiger92, Zhang98, MG2000, Gieseke2000, Bass03, Xu05, Greco05,
Bleicher99}. This approach is applicable for investigations of such phenomena as thermalization,
kinetic decoupling, and dynamics of high-energy particles in systems far from equilibrium, 
i.e., in a regime where the second-order viscous hydrodynamics breaks down \cite{HM08}.

Recently, an on-shell parton cascade Boltzmann Approach of MultiParton
Scatterings (BAMPS) has been developed to study 
thermalization \cite{Xu05,Xu07,El08}, elliptic flow $v_2$ \cite{XGSPRL08,XG08v2,XGS08},
and the energy loss \cite{FXG08} of gluons produced in Au+Au collisions
at RHIC energy. Also the generation and evolution of viscous shock 
waves are surprisingly well realized in BAMPS calculations \cite{Bouras}.
The shear viscosity of the gluon matter at RHIC has been estimated from 
BAMPS calculations \cite{XG08v2, XGS08} within the Navier-Stokes 
approximation \cite{XUPRL}. The authors found that to produce large 
$v_2$ comparable with the experimental data, the gluon matter
should have an $\eta/s$ between $0.08$ and $0.2$ constrained by
details of the hadronization and the kinetic freeze out. This is in line with the dissipative hydrodynamic approach \cite{Romatschke08}. 
Perturbative QCD (pQCD) gluon bremsstrahlung $gg\leftrightarrow ggg$
is responsible for the low $\eta/s$ ratio and for the generation
of large elliptic flow.

Beyond the Navier-Stokes approximation, which has been used in Refs.\cite{AYM,XUPRL}, we derive a new microscopic formula for the shear viscosity coefficient from the kinetic theory using the second-order Grad's method. This is one of the
goals in the present article. The derivation follows Ref. \cite{AM107} and is generalized for a particle system out of chemical equilibrium. 

Another goal is to elaborate on the breakdown region of the second-order
viscous hydrodynamics. To do this we investigate the time evolution of 
a gluon matter in a one-dimensional expansion with Bjorken boost 
invariance \cite{Bjorken83} by solving the Israel-Stewart hydrodynamic 
equations \cite{HM08} as well as by performing similar BAMPS transport calculations for comparison.
We quantify the deviation of the gluon distribution function from kinetic
equilibrium and show the region with large deviation, where the
applicability of the Israel-Stewart hydrodynamics is questionable.

The article is organized as follows. In Sec. \ref{tk} we introduce 
theoretical framework for deriving viscosity from the kinetic 
theory using second-order Grad's method. We consider a massless particle system, which undergoes a one-dimensional expansion with Bjorken boost invariance.  A comparison with the results obtained by the Navier-Stokes approximation \cite{XUPRL} is given in Sec. \ref{comparison}. Using the formula derived in Sec. \ref{tk} 
we calculate the shear viscosity to entropy density ratio $\eta/s$ of 
gluon matter: in Sec. \ref{iterative_method} an iterative and 
self-consistent approach is introduced to calculate $\eta/s$ from the 
Israel-Stewart hydrodynamics, whereas the results from BAMPS calculations
are presented in Sec. \ref{results}. For both hydrodynamic and transport calculations, deviations from kinetic as well as
chemical equilibrium are shown and analyzed.  Conclusions are given in 
Sec. \ref{conclusions}. 

\section{Shear viscosity coefficient from second-order kinetic theory}
\label{tk}
Relativistic causal dissipative  hydrodynamic equations can be derived from the kinetic theory by applying Grad's method of moments \cite{Grad49}. A detailed derivation is reported in Refs. \cite{AM107,DHR93} and a prescription for calculating transport coefficients is also given there. In this section we will follow Ref. \cite{AM107} to derive an expression for the shear viscosity coefficient $\eta$ when the considered system is out of chemical equilibrium.

The basic equation of relativistic kinetic theory is the Boltzmann equation
\begin{equation}
p^{\mu} \partial_{\mu} f(x,p)=C[f(x,p)]
\label{be}
\end{equation}
for a one-particle phase-space distribution function 
$f(x,p))=\frac{dN}{\frac{1}{(2\pi)^3}\,d^3 p \,d^3 x}$.
$C[f(x,p)]$ denotes the collision term, which accounts for all microscopic interaction processes among particles.
The entropy four-current is defined by \cite{AM107,deGroot}
\begin{equation}
s^{\mu}=-\int \frac{d^3 p}{(2\pi)^3 p^0} p^{\mu} f(x,p) \left[ \ln (f(x,p))-1 \right] .
\label{s_kinetic}
\end{equation}
The entropy production is then given by
\begin{equation}
\partial _{\mu}s^{\mu}=-\int dw \, p^{\mu}\partial_{\mu}f(x,p)\ln f(x,p)
=-\int dw \, C[f(x,p)] \ln f(x,p)
\label{ds}
\end{equation}
with the short notation $dw=\frac{d^3 p}{(2\pi)^3 p^0}$. 

We now assume that the deviation of $f(x,p)$ from the equilibrium distribution $f_{eq}(x,p)$ is small:
\begin{equation}
f(x,p)=f_{eq}(x,p)\left(1+\phi(x,p) \right)
\label{f_expansion}
\end{equation}
where $\phi(x,p)\ll 1$ and
\begin{equation}
\label{feq}
f_{eq}(x,p)=\lambda \, e^{-\frac{u_{\mu}p^{\mu}}{T}}\,.
\end{equation}
$\lambda(x)$ and $T(x)$ denote the local fugacity and temperature, respectively. $u^{\mu}(x)$ is the hydrodynamic four-velocity of the medium. Equation (\ref{feq}) is the standard form for Boltzmann particles. The derivation below can be
easily extended for Bose and Fermi particles. In addition, we will restrict the following discussions to the case of massless particles (e.g., gluons).

We expand $\phi(x,p)$ up to second order in momentum, that is,
\begin{equation}
\phi(x,p) = \epsilon(x)-\epsilon_{\mu}(x)p^{\mu}+\epsilon_{\mu\nu}(x)p^{\mu}p^{\nu} \,,
\label{phixp}
\end{equation} 
where the momentum-independent coefficients can be expressed in terms of the dissipative currents $\Pi$, $q^\mu$ and $\pi^{\mu\nu}$ denoting bulk pressure, heat flux and shear tensor \cite{AM107,DHR93}: 
\begin{eqnarray}
\label{epsilonmv}
\epsilon_{\mu\nu} &=& A_2(3u_\mu u_\nu-\Delta_{\mu\nu})\Pi-B_1 u_{(\mu}q_{\nu)}+C_0\pimn \\ 
\epsilon_{\mu} &=& A_1 u_\mu \Pi-B_0 q_{\mu} \\
\epsilon &=& A_0\Pi 
\label{epsilons}
\end{eqnarray}
with the projector $\Delta^{\mu\nu}=g^{\mu\nu}-u^\mu u^\nu$ and symmetrization operation $u_{(\mu}q_{\nu)}=\frac{1}{2}(u_\mu q_\nu+u_\nu q_\mu)$. The metric used in this work is $g^{\mu\nu}=diag(1,-1,-1,-1)$.
In general, the dissipative fluxes are defined as projections of deviations of the energy-momentum tensor $T^{\mu\nu}$ and particle four-current $N^\mu$ from their equilibrium form  \cite{AM107,DHR93}:
\begin{eqnarray}
\label{bulk}
\Pi &=& -\frac{1}{3}\Delta_{\mu\nu}\delta T^{\mu\nu} \\
\label{heatf}
q^{\mu} &=& \Delta_{\nu}^{\mu}u_{\rho}\delta T^{\rho\nu} - \frac{4}{3}\Delta_{\nu}^{\mu}\delta N^{\nu} \\
\label{diss_fluxes}
\pi^{\mu\nu} &=& \delta T^{<\mu\nu>} = \left( \frac{1}{2}\Delta_{\alpha}^{\mu}\Delta_{\beta}^{\nu} + \frac{1}{2}\Delta_{\alpha}^{\nu}\Delta_{\beta}^{\mu}- \frac{1}{3}\Delta_{\alpha\beta}\Delta^{\mu\nu}\right) \delta T^{\alpha\beta}
\end{eqnarray}
with the definitions $N^\mu=\int dw p^\mu f$ , $T^{\mu\nu}=\int dw p^\mu p^\nu f$ and  $\delta T^{\mu\nu}=T^{\mu\nu}-T_{eq}^{\mu\nu}$, $\delta N^{\mu}=N^{\mu}-N^{\mu}_{eq}$. 

We use the following local matching conditions on the energy and particle densities:
\begin{eqnarray}
\label{matche}
e=e_{eq}=\frac{3\lambda T^4 }{\pi^2} \\
\label{matchn}
n=n_{eq}= \frac{\lambda T^3}{\pi^2}
\end{eqnarray}
with the definitions for the densities $e=u_\mu T^{\mu\nu} u_\nu$ and $n=u_\mu N^\mu$. The local temperature simply follows as $T=e/3n$. The fugacity is then calculated via $\lambda=n/(\frac{1}{\pi^2}T^3)$.
One obtains immediately $u_\mu \delta T^{\mu\nu} u_\nu=0$ and $u_\mu \delta N^\mu=0$. The bulk pressure $\Pi$ from Eq. (\ref{bulk}) then becomes
\begin{equation}
\Pi\sim (g_{\mu\nu}-u_{\mu} u_{\nu})\delta T^{\mu\nu}=\delta T^{\nu}_{\nu}=0 
\end{equation}
for massless particles, since the energy momentum tensor is traceless in this case. Thus, $\epsilon=0$ according to Eq. (\ref{epsilons}).

In the following, we will consider a one-dimensional Bjorken expansion \cite{Bjorken83}. This implies that in the local rest frame, the distribution function $f(x,p)$ is symmetric when transforming $\vec p$ to $-\vec p$. Thus in the local rest frame, $T^{0i}=0$ and $N^i=0$, where $i=1,2,3$. 
The heat flux $q^\mu$ (\ref{heatf}) vanishes in the local rest frame because
\begin{equation}
q^\mu=g_\nu^\mu u_\rho \delta T^{\rho\nu}-u^\mu u_\nu\delta T^{\rho\nu}-\frac{4}{3}g_\nu^\mu \delta N^\nu+\frac{4}{3}u^\mu u_\nu \delta N^\nu=u_\rho \delta T^{\rho\mu}-\frac{4}{3}\delta N^\mu=0 \, .
\label{heat_flux}
\end{equation}
We obtain then $\epsilon_\mu p^\mu \sim q_\mu p^\mu = 0$ [see Eq. (\ref{epsilons})]. 

For a one-dimensionally expanding system, Eq. (\ref{phixp}) thus reduces to
\begin{equation}
\phi(x,p) = \epsilon_{\mu\nu}(x)p^{\mu}p^{\nu} \,.
\label{phixp_redu}
\end{equation}

Putting $f=f_{eq}(1+\phi)$ into Eq. (\ref{ds}) and using the linearization
\begin{equation}
 \ln(1+\phi) \approx \phi=\epsilon_{\mu\nu}(x) p^\mu p^\nu
\end{equation}
we rewrite Eq.(\ref{ds}) as
\begin{equation}
\partial _{\mu}s^{\mu} = -\int dw \, C[f(x,p)]\ln f_{eq}(x,p)-\int dw \, C[f(x,p)]\epsilon_{\mu\nu}p^{\mu}p^{\nu}
\label{ds2}
\end{equation}
Using the formula (\ref{feq}) for $f_{eq}$ in the first term of Eq. (\ref{ds2})
one has
\begin{eqnarray}
&& -\int dw \, C[f(x,p)]\,\ln f_{eq}(x,p) = -\int dw C[f(x,p)]\, (\ln \lambda-u_{\mu}p^{\mu}/T) \nonumber \\
&=& -\ln \lambda \int dw \, C[f(x,p)]+u_\mu \int dw \, p^\mu \, C[f(x,p)]/T \nonumber \\
&=& -\ln \lambda \int dw \, C[f(x,p)] = -\ln \lambda \,\partial_{\mu}N^{\mu} \,.
\label{intCf}
\end{eqnarray}
For the second-last identity in Eq. (\ref{intCf}), we used the energy-momentum conservation: $\int dw \, p^\nu \, C[f(x,p)]=\partial_\mu \int dw \, p^\nu p^\mu f=\partial_\mu T^{\nu \mu}=0$. Equation (\ref{intCf}) describes entropy production due to particle production ($\partial_{\mu}N^{\mu}>0$ for $\lambda < 1$) and absorption ($\partial_{\mu}N^{\mu}<0$ for $\lambda > 1$).

With the definitions 
\begin{eqnarray}
\label{pmunu}
P^{\mu\nu} &=& \int dw p^{\mu} p^{\nu} C[f(x,p)] \\
\label{cbar}
{\bar C} &=& \int dw C[f(x,p)]=\partial_{\mu}N^{\mu}\,,
\end{eqnarray}
which are the 2nd and the 0th moment of the collision term the entropy production in Eq.(\ref{ds2}) can be now written in a more compact form
\begin{equation}
\partial_{\mu}s^{\mu}=-{\bar C}\ln \lambda - \epsilon_{\mu\nu}P^{\mu\nu}\,.
\label{ds3a}
\end{equation}
In general, the entropy production in an imperfect fluid can be expressed by the positive definite form  \cite{IS,Elze01,AM04}
\begin{equation}
\partial_{\mu}s^{\mu}=-J \ln \lambda+(\zeta T)^{-1}\Pi^2-(\kappa T)^{-1}q_{\alpha}q^{\alpha}+(2\eta T)^{-1}\pi_{\alpha\beta}\pi^{\alpha\beta}\,,
\label{ds3b}
\end{equation}
where $\zeta$, $\kappa$, and $\eta$ are non-negative coefficients denoting the bulk viscosity, heat conductivity and shear viscosity, respectively. 
$J=\partial_{\mu}N^{\mu}$ is the source of particle production \cite{Elze01,AM04} and is identical with $\bar C$ (\ref{cbar}). For a chemically equilibrated system $J$ vanishes. 
Comparing Eq. (\ref{ds3a}) to (\ref{ds3b}) we find
\begin{equation}
- \epsilon_{\mu\nu}P^{\mu\nu} = (2\eta T)^{-1} \pi_{\alpha\beta}\pi^{\alpha\beta}\,,
\label{ds3}
\end{equation}
because in our case $\Pi=0$ and $q_\alpha q^\alpha=0$ as discussed above.
The expression (\ref{ds3}) is exactly the same as obtained in \cite{AM107} and describes entropy production due to shear viscous effects.

We then obtain the final expression for the shear viscosity coefficient
\begin{equation}
\eta = -\frac{ \pi_{\alpha\beta}\pi^{\alpha\beta}}{2 T \epsilon_{\mu\nu}P^{\mu\nu}}
=-\frac{ \pi_{\alpha\beta}\pi^{\alpha\beta}}{2 T  C_0 \pi_{\mu\nu}P^{\mu\nu}}\,.
\label{eta}
\end{equation}
The last identity is due to the fact that $q^\mu$ vanishes in the local
rest frame and thus $u_{(\mu}q_{\nu)} P^{\mu\nu}=0$. We note that the derived formula (\ref{eta}) is an approximate expression of the true shear viscosity. We call the ``second-order'' shear viscosity, because we have used terms up to second order in momentum for $\phi(x,p)$ [see Eq. (\ref{phixp})]. 

To calculate $C_0$ we go to the local rest frame, i.e., 
$u^{\mu}=(1,0,0,0)$, where
\begin{equation}
\pi^{\mu\nu}=\delta T^{\mu\nu}= T^{\mu\nu}-T_{eq}^{\mu\nu}=\epsilon_{\alpha\beta}\int dw \, p^{\mu}p^{\nu}p^{\alpha}p^{\beta} f_{eq}(x,p)
\label{pimunu}
\end{equation}
is valid according to Eqs. (\ref{diss_fluxes}) and (\ref{phixp_redu}) for a (0+1) dimensional expansion.
In this frame $\epsilon_{\alpha\beta}$ [see Eq. (\ref{epsilonmv})] reduces to
\begin{equation}
\epsilon_{\alpha\beta} = C_0 \pi_{\alpha\beta}\,.
\label{epsilonmunu2}
\end{equation}
Calculating the integrals in Eq. (\ref{pimunu}) with $f_{eq}=\lambda e^{-E/T}$
gives
\begin{eqnarray}
&& (1-C_0 40 \lambda T^6/\pi^2) \pi^{0j}=0\,,\quad j=1,2,3\\
&& (1-C_0 8 \lambda T^6/\pi^2) \pi^{ij}=0\,,\quad i,j=1,2,3\,.
\end{eqnarray}
We have used the fact that $\pi^{\mu\nu}$ is traceless and $\pi^{00}=0$
due to the matching condition (\ref{matche}) and $T^{00}=e$ in the local
rest frame. For a system undergoing a one-dimensional Bjorken
expansion, i.e., in a (0+1) dimensional case, all off-diagonal elements of $T^{\mu\nu}$ - and thus 
$\pi^{\mu\nu}$ as well - vanish in the local rest frame, particularly $T^{0j}=\pi^{0j}=0, j=1,2,3$. Thus we obtain
\begin{equation}
 C_0=\frac{\pi^2}{ 8 \lambda T^6} \,.
\label{c_0}
\end{equation} 

If the third spatial coordinate is chosen as the expansion axis, we have $T^{11}=T^{22}$, and in the local rest frame the shear tensor takes the form
\begin{equation}
\label{pimunu_matrix}
\pi^{\mu\nu}=
\begin{pmatrix} 
0 & 0 & 0 & 0 \\ 0 & -\frac{{\bpi}}{2} & 0 & 0 \\ 0 & 0 & -\frac{{\bpi}}{2} & 0 \\ 0 & 0 & 0 & \bpi
\end{pmatrix}
\end{equation} 
which is also given in \cite{AM04}. We thus obtain
\begin{eqnarray}
\label{pimunu2}
&& \pi_{\mu\nu}\pi^{\mu\nu}=\frac{3}{2}{\bar \pi}^2 \\
\label{epmunu}
&& \epsilon_{\mu\nu}P^{\mu\nu}=C_0\pi_{\mu\nu}P^{\mu\nu}
=\frac{C_0{\bar \pi}}{2}(3P^{33}-P^{00}) \,,
\end{eqnarray} 
where we have used $P^{11}+P^{22}=P^{00}-P^{33}$, because $P^{\mu\nu}$
is traceless following from the definition (\ref{pmunu}). Putting Eqs. (\ref{pimunu}) and (\ref{epmunu}) into (\ref{eta})
gives the shear viscosity coefficient for a (0+1) dimensionally  expanding system of 
massless particles:
\begin{equation}
\eta=-\frac{3\bpi}{2 T C_0 (3P^{33}-P^{00})}= 
4n\frac{-T^2\bpi}{P^{33}-\frac{1}{3}P^{00}} \,.
\label{etacompact}
\end{equation}
For the last identity, we have used the matching conditions (\ref{matchn}) and Eq. (\ref{c_0}).

The energy density $e$, the temperature $T$ and the shear component $\bpi$
in a (0+1) dimensional expansion
can be calculated by solving viscous hydrodynamic equations with a given value of shear
viscosity $\eta$. If $\bpi$ is known, the distribution function $f$ is
known too [see Eqs. (\ref{f_expansion}), (\ref{phixp_redu}) and
(\ref{epsilonmunu2})]. One can thus evaluate $P^{00}$ and $P^{33}$
according to their definitions (\ref{pmunu}). Then $\eta$ can be
calculated employing Eq. (\ref{etacompact}). 
In sSec. \ref{iterative_method} we will introduce an iterative and
self-consistent prescription to calculate the second-order shear viscosity.

On the other hand, $f$ can be obtained by solving the
Boltzmann equation (\ref{be}) directly employing transport simulations.
Then $\eta$ can be easily extracted using Eq. (\ref{etacompact}).
Such calculations will be presented in section \ref{results}. The results
will be compared with those obtained in Sec. \ref{iterative_method}.
As it turns out, a ratio of mean transport free path to expansion time being larger than unity and the variance of $\phi(x,p)$ being larger than unity
will possibly indicate the breakdown of the second-order viscous
hydrodynamics. In this regime the validity of (\ref{etacompact}) is 
also questionable.

\section{Comparison to shear viscosity from Navier-Stokes approximation}
\label{comparison}
In Ref. \cite{XUPRL}, the shear viscosity coefficient was derived assuming
the Navier-Stokes approximation
\begin{equation}
\label{NS}
\pi^{\mu\nu}=2\eta \nabla^{<\mu}u^{\nu>} \,.
\end{equation}
It reads
\begin{equation}
\label{shv2}
\eta_{NS} \cong \frac{1}{5} n \frac{\langle E/3-p_z^2/E) \rangle}
{\frac{1}{3}-\langle p_z^2/E^2 \rangle} \frac{1}{\sum R^{\rm tr}+ 
\frac{3}{4} \partial_t (\ln \lambda)}
\end{equation}
where 
\begin{equation}
\label{trate}
\sum R^{\rm tr}= \frac{\int dw\, \frac{p_z^2}{E^2} \, C[f] -
\langle p_z^2/E^2 \rangle \int dw\, C[f]}{n\, (\frac{1}{3}-
\langle p_z^2/E^2 \rangle)}
\end{equation}
is the total transport collision rate, which was introduced in \cite{Xu07}.
All integrals are expressed in the local rest frame. $\langle \rangle$ denotes the average over particle momentum.

Equation (\ref{eta}) can be used to calculate the shear viscosity if the shear tensor $\pimn$ obeys the Israel-Stewart equation
\cite{AM04}
\begin{equation}
\tau_\pi \Delta_\mu^\alpha \Delta_\nu^\beta \dot \pi_{\alpha\beta}+\pimn=2\eta\sigma_{\mu\nu}-\big[ \eta T\partial_\lambda\left( \frac{\tau_\pi}{2\eta T}u^\lambda \right)\pimn \big] \,,
\label{IS}
\end{equation}
where $\sigma_{\mu\nu}=\nabla^{<\mu}u^{\nu>}$ and $\tau_\pi$ denotes 
the relaxation time [see also Eq. (\ref{2nd_order1}) below]. Equation (\ref{IS}) is more general than (\ref{NS}) in the first-order (Navier-Stokes) theory.

If we define
\begin{equation}
\label{Rtr_IR}
\sum R^{tr}_{Grad}=\frac{P^{33}-\frac{1}{3}P^{00}}{n\left( \frac{1}{3} \langle E^2\rangle-\langle p_z^2\rangle\right)} \,,
\end{equation}
then the shear viscosity from the Grad's method (\ref{etacompact}) can be rewritten to
\begin{equation}
\label{etais}
\eta_{Grad}=4n \frac{T^2\langle E/3 - p_z^2/E \rangle }{\frac{1}{3} \langle E^2\rangle-\langle p_z^2\rangle} \, \frac{1 }{\sum R^{tr}_{Grad}}\,,
\end{equation} 
where we have used 
$\bpi=T^{33}-T^{33}_{eq}=T^{33}-\frac{1}{3}T^{00}=n\langle p_z^2/E-E/3 \rangle $. 
Remember that $P^{\mu\nu}$ is the second moment of the collision term [see Eq. (\ref{pmunu})].
The expression (\ref{etais}) is similar to Eq. (\ref{shv2}) except
for the term $\frac{3}{4} \partial_t (\ln \lambda)$, which indicates that chemical equilibration contributes explicitly
to the shear viscosity in the Navier-Stokes approximation rather than in the Israel-Stewart approach.

In the next section, we calculate the shear viscosity in a gluon system
within the Israel-Stewart approach and compare the result with that
obtained using the Navier-Stokes approximation \cite{XUPRL}.

\section{Calculation of shear viscosity in a gluon system: an iterative
and self-consistent prescription}
\label{iterative_method}
In this section we want to calculate the shear viscosity to the entropy density ratio $\eta/s$ for a gluonic system, which undergoes a one-dimensional expansion with Bjorken boost invariance, i.e., a (0+1) dimensional expansion.

\subsection{Prescription}
For a (0+1) dimensional case the shear tensor $\pi_{\mu\nu}$ in the
local rest frame is given by Eq. (\ref{pimunu_matrix}). Then the gluon distribution function in the local rest frame reads  
\begin{equation}
f(x,p) = \lambda e^{-\frac{E}{T}} \left[ 1 - C_0\bpi (p_z^2 - p_t^2/2) \right ]
\label{fxp2} 
\end{equation}
according to Eqs. (\ref{f_expansion}), (\ref{phixp_redu}), (\ref{epsilonmunu2}) and (\ref{pimunu_matrix}).
If $\bpi$, $T$ and $\lambda$ are known, the shear viscosity $\eta$ can be calculated according to Eq. (\ref{etacompact}), where $P^{\mu\nu}$ are evaluated by Eq. (\ref{pmunu}) via Eq. (\ref{fxp2}). Note that for the case of a gluonic system the value of $\eta$ has to be amplified by the degeneracy factor of gluons $d_G=16$.
We thus define $\eta_g=d_G \eta$. In addition, the gluon entropy density is given by
\begin{equation}
s_g = u_\mu s^\mu= -d_G \int dw p_0 f(x,p) (\ln f(x,p)-1)\approx (4-\ln \lambda)\, n_g
-\frac{9\bpi_g^2}{8 n_g T^2}\,,
\label{sdens}
\end{equation}
where $n_g=d_G \lambda T^3/\pi^2$ and $\bpi_g=d_G \bpi$ are the gluon number 
density and the gluon shear component, and we have used the approximation 
$\ln(1+\phi)\approx \phi$ for small $\phi=-C_0 \bpi (p_z^2 - p_t^2/2)$.
We note that $\phi$ can be larger than unity for large momenta. In these cases,
the expansion [also for Eq. (\ref{ds2})] fails. On the other hand, the 
distribution function $f(x,p)$ becomes very small at large momenta. The effect
of the invalid expansion on the integrated quantity $s_g$ is thus 
negligible at this point. 

In principle, $\phi=(f-f_{eq})/f_{eq}$ gives the relative deviation from 
kinetic equilibrium. However, $\phi$ is also a function of momentum. 
The average $\langle \phi(x,p) \rangle_{eq}$ over momentum distributed in
equilibrium, i.e, using $f(x,p)$ in zeroth order of $\bpi$, is obviously zero.
We introduce the variance $\sigma_\phi=\sqrt{\langle \phi^2 \rangle_{eq}}$
as the quantity determining the deviation from kinetic equilibrium and
we find
\begin{equation}
\label{sigmaphi}
\sigma_\phi=\frac{9\sqrt{2}}{4} \, \frac{|\bpi_g|}{e_g}\,,
\end{equation}
where $e_g=3n_gT$ is the gluon energy density.

If the deviation from the local kinetic equilibrium is sufficiently small,
then the dynamical expansion in a (0+1) dimensional case can be well described
by the Israel-Stewart (IS) viscous hydrodynamic equations \cite{IS,DHR93,AM107,AM04,BRW06,HM08}:
\begin{eqnarray}
\label{2nd_order_n}
\frac{dn_g}{d\tau} &=& -\frac{n_g}{\tau} \,,\\
\label{2nd_order_e}
\frac{de_g}{d\tau} &=& -\frac{4}{3}\frac{e_g}{\tau}+\frac{\bpi_g}{\tau} \,,\\
\frac{d\bpi_g}{d\tau} &=& -\frac{\bpi_g}{\tau_{\pi}}-\frac{1}{2}\bpi_g \left( \frac{1}{\tau} + \frac{1}{\beta_2}T\frac{\partial}{\partial\tau}(\frac{\beta_2}{T}) \right) + \frac{2}{3}\frac{1}{\beta_2\tau} \,,
\label{2nd_order1}
\end{eqnarray}
where $\beta_2=9/(4e_g)$ and $\tau_{\pi}=2\beta_2\eta_g$ denotes 
the relaxation time. Equation (\ref{2nd_order1}) is just Eq. (\ref{IS}) expressed in the local rest frame using the 
hydrodynamic velocity $u^\mu=\frac{1}{\tau}(t,0,0,z)$, where 
$\tau=\sqrt{t^2-z^2}$. In derivation of Eq. (\ref{IS}), which is discussed in Ref. \cite{AM04}, only terms of second order in gradients and dissipative flux $\pimn$ have been included. If $\sigma_\phi$ in Eq. (\ref{sigmaphi}) is larger than unity, further terms containing $\sigma_\phi^2\sim (\bpi/e)^2\sim (\beta_2\bpi)^2$ are no longer small enough anymore to be ommited in derivation of Eq. (\ref{IS}) and thus in Eq. (\ref{2nd_order1}) as well, i.e., a higher order hydrodynamic equation is needed. Thus the value of $\sigma_\phi$ is an indicator for a breakdown of second-order hydrodynamic theory. 

Equation (\ref{2nd_order_n}) for the gluon density can be easily solved:
\begin{equation}
n_g(\tau)=n_g(\tau_0) \, \frac{\tau_0}{\tau} \,,
\end{equation}
which is identical with the result from ideal hydrodynamics.
On the other hand, the energy density decreases slower than in 
ideal hydrodynamics due to the viscous effects:
\begin{equation}
e_g(\tau)=e_g(\tau_0) \left (\frac{\tau_0}{\tau} \right )^\xi\,, \quad \xi \le 
\frac{4}{3} \,.
\end{equation}
Thus we obtain the gluon fugacity
\begin{equation}
\label{fuga_quali}
\lambda(\tau)=\frac{n_g(\tau)}{n_g^{eq}(\tau)}= 
\frac{n_g}{\frac{d_G}{\pi^2}T^3}=
\frac{n_g}{\frac{d_G}{\pi^2} (e_g/3n_g)^3}=
\lambda_0 \left (\frac{\tau_0}{\tau} \right )^{4-3\xi} \le \lambda_0 \,,
\end{equation}
where $\lambda_0=\lambda(\tau_0)$.
The system will be continuously out of chemical equilibrium during 
the expansion, even if it is initially at local thermal equilibrium 
($\lambda_0=1$). The larger the viscosity, the smaller is the value of $\xi$
and the faster is the decrease of the fugacity. Inclusion of 
production and annihilation processes such as the gluon bremsstrahlung
and its back reaction ($gg\leftrightarrow ggg$) makes chemical
equilibration possible and thus, of course, Eq. (\ref{2nd_order_n}) has to be modified! 
However, in this work we will use Eq.(\ref{2nd_order_n}) without any modifications. 
The derivation of new and altered equations and their solutions will be given in a forthcoming 
publication \cite{el2}.

One can solve Eqs. (\ref{2nd_order_e}) and (\ref{2nd_order1}), if
the initial values of $n_g$, $e_g$, $\bpi_g$ and also the value of the shear
viscosity $\eta_g$ are given. On the other hand, to calculate $\eta_g$ using Eq. 
(\ref{etacompact}) via Eq. (\ref{fxp2}) we need $n_g$, $e_g$, and $\bpi_g$. It is
obvious that an iterative algorithm has to be developed to calculate
$n_g$, $e_g$, $\bpi_g$ and $\eta_g$ self-consistently. This algorithm is as follows:
\begin{enumerate}
\item
We solve Eqs. (\ref{2nd_order_n})-(\ref{2nd_order1})
with a guessed value of $\eta_g$. The guessed value can be chosen arbitrarily because the final result does not depend on it. $\eta_g/n_g$ is assumed to be a constant of time.
\item
The obtained $n_g(\tau)$, $e_g(\tau)$ and $\bpi_g(\tau)$ at a time $\tau$ 
are used to
calculate $\eta_g(\tau)$ according to (\ref{etacompact}). We calculate first the moments $P^{\mu\nu}$ using $f(x,p)$ in Eq. (\ref{fxp2}) with given $n_g(\tau), e_g(\tau)$ and $\bpi_g(\tau)$.
\item
We turn back to step 1. The value of $\eta_g(\tau)$ is used to solve 
Eqs. (\ref{2nd_order_e}) and (\ref{2nd_order1}) again. 
\end{enumerate}
Iterations will continue, until the relative deviation of $\eta_g$
from the previous one is sufficient small.  The iterative procedure allows to calculate $\bpi(\tau)$, $e(\tau)$ and $n(\tau)$ as well as $\eta/s(\tau)$ in a consistent way for given interactions. We note that if $\eta_g/n_g$ is
strongly time dependent, further iterations will be required to account for this time dependence. A refined
algorithm will be presented in \cite{el2}.

To obtain $\eta_g$, $P^{\mu\nu}$ has to be first evaluated by (\ref{pmunu}) 
via (\ref{fxp2}). $P^{\mu\nu}$ is a second moment of the collision term and thus is determined by gluon interactions considered. The compact forms of the collision terms can be found in \cite{Xu05}. In this article elastic ($gg\to gg$) as well as bremsstrahlung 
($gg\leftrightarrow ggg$) processes inspired within perturbative QCD are responsible
for the gluon dynamics. The differential cross section and the effective 
matrix element are taken as in Refs. \cite{Xu05,El08}:
\begin{eqnarray}
\label{cs22}
\frac{d\sigma^{gg\to gg}}{dq_{\perp}^2} &=&
\frac{9\pi\alpha_s^2}{(q_{\perp}^2+m_D^2)^2} \,, \\
\label{m23}
| {\cal M}_{gg \to ggg} |^2 &=&\frac{9 g^4}{2}
\frac{s^2}{({\bf q}_{\perp}^2+m_D^2)^2}\,
 \frac{12 g^2 {\bf q}_{\perp}^2}
{{\bf k}_{\perp}^2 [({\bf k}_{\perp}-{\bf q}_{\perp})^2+m_D^2]}\,
\Theta(k_{\perp}\Lambda_g-\cosh y)
\label{matrix}
\end{eqnarray}
where $g^2=4\pi\alpha_s$. The Debye screening mass 
\begin{equation}
\label{md}
m_D^2=d_G \pi \alpha_s \int dw \, N_c \, f(x,p)
\end{equation}
with $N_c=3$ is applied to regularize infrared divergences. 
Although $gg\leftrightarrow ggg$ processes are considered, they contribute
only to the shear viscosity but not to chemical equilibration, because
as mentioned above, particle number conservation is assumed at present to derive
Eq. (\ref{2nd_order_n})). Improvements will be done in a forthcoming publication \cite{el2}.

\subsection{Results}
\begin{figure}[p]
\epsfbox{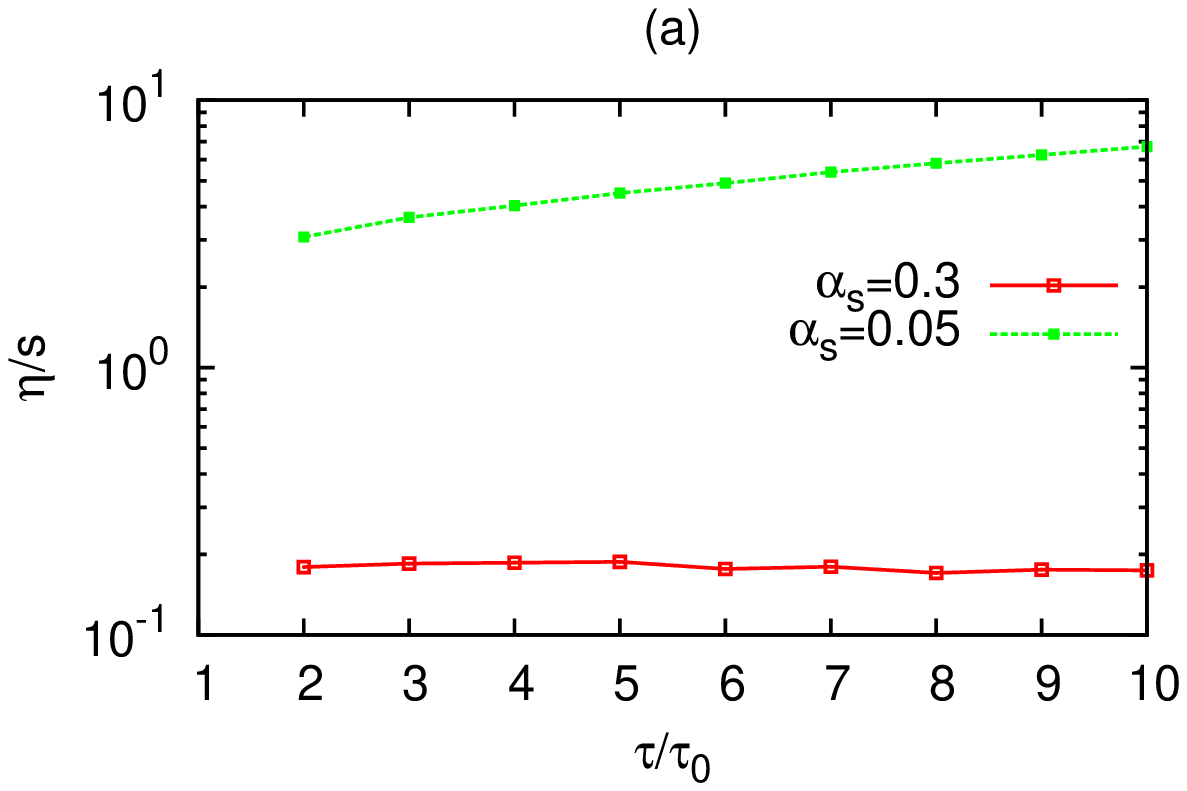}
\vfill
\epsfbox{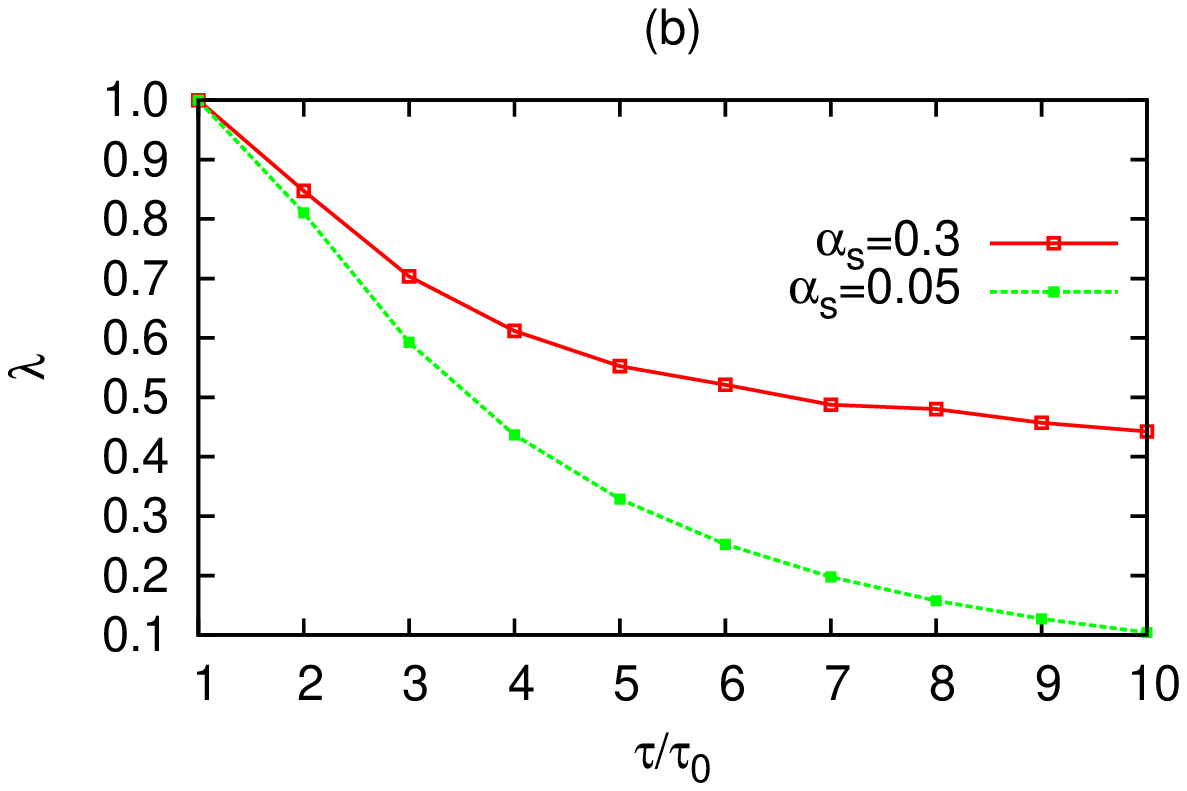}
\caption {(Color online)  (a) $\eta/s$ ratio and (b)  fugacity $\lambda$ calculated by the iterative procedure described in the text for $\alpha_s=0.05$ and $\as=0.3$ at ten different time points, with initial time $\tau_0=0.4$ fm/c, $T(\tau_0)=500$ MeV. The initial input value of $\eta/s$ is $0.5$.}
\label{eos_iter}
\end{figure}

Figure \ref{eos_iter}(a) shows $\eta_g/s_g$ as a function of
the expansion time for two values of the coupling constant $\alpha_s=0.05$ and $0.3$.
The initial gluon system at $\tau_0=0.4$ fm/c is assumed to be in thermal
equilibrium with a temperature of $T_0=500$ MeV. Each of the results
indicated by the symbols in Fig. \ref{eos_iter} is obtained by about 40
iterations with a guessed value of $\eta_g({\rm guessed})=0.5 \, s_g^{eq}$.
From Fig. \ref{eos_iter}(a) we see that the ratio $\eta_g/s_g$ is almost constant in
time for $\alpha_s=0.3$, whereas for $\alpha_s=0.05$, $\eta_g/s_g$ increases
moderately. The assumption underlying the iterative algorithm that
$\eta_g/n_g \approx 4\eta_g/s_g$ does not depend on time is justified accordingly. 
One finds that $\eta_g/s_g\approx 0.18$ for a coupling of $\alpha_s=0.3$ and $\eta_g/s_g\approx 3$ for $\alpha_s=0.05$.

The results for the gluon fugacity (obtained from the solution of Eqs.(\ref{2nd_order_n})-(\ref{2nd_order1})) depicted in 
Fig. \ref{eos_iter}(b) show a strong time dependence. The smaller the value 
of $\alpha_s$, i.e., the larger the $\eta_g/s_g$, the faster is the 
deviation from the chemical equilibrium. This quantitatively demonstrates the 
consideration from above [see Eqs. (\ref{fuga_quali})].

When putting Eq. (\ref{fxp2}) into Eq. (\ref{pmunu}) one realizes that 
$P^{\mu\nu}\sim \lambda^2 C_0 \bpi\sim \lambda\bpi$ in leading order of $\bpi$. 
Thus $\eta_g$ does not depend on $\lambda$. Secondly, from Eq. (\ref{sdens}) we obtain 
$s_g/T^3 \sim \lambda (1-\ln \lambda )$. Thus, 
$\eta_g/s_g \sim 1/\lambda(1-\ln \lambda)$ and will increase slower than a logarithmical behavior
when $\lambda$ decreases: a stronger decrease of $\lambda$ (comparing the 
result for $\alpha_s=0.05$ with that for $\alpha_s=0.3$ in the lower panel of
Fig. \ref{eos_iter}) will lead to stronger increase 
of $\eta_g/s_g$, as seen in the numerical results shown in Fig. \ref{eos_iter}(a).
 
\begin{figure}[p]
\epsfbox{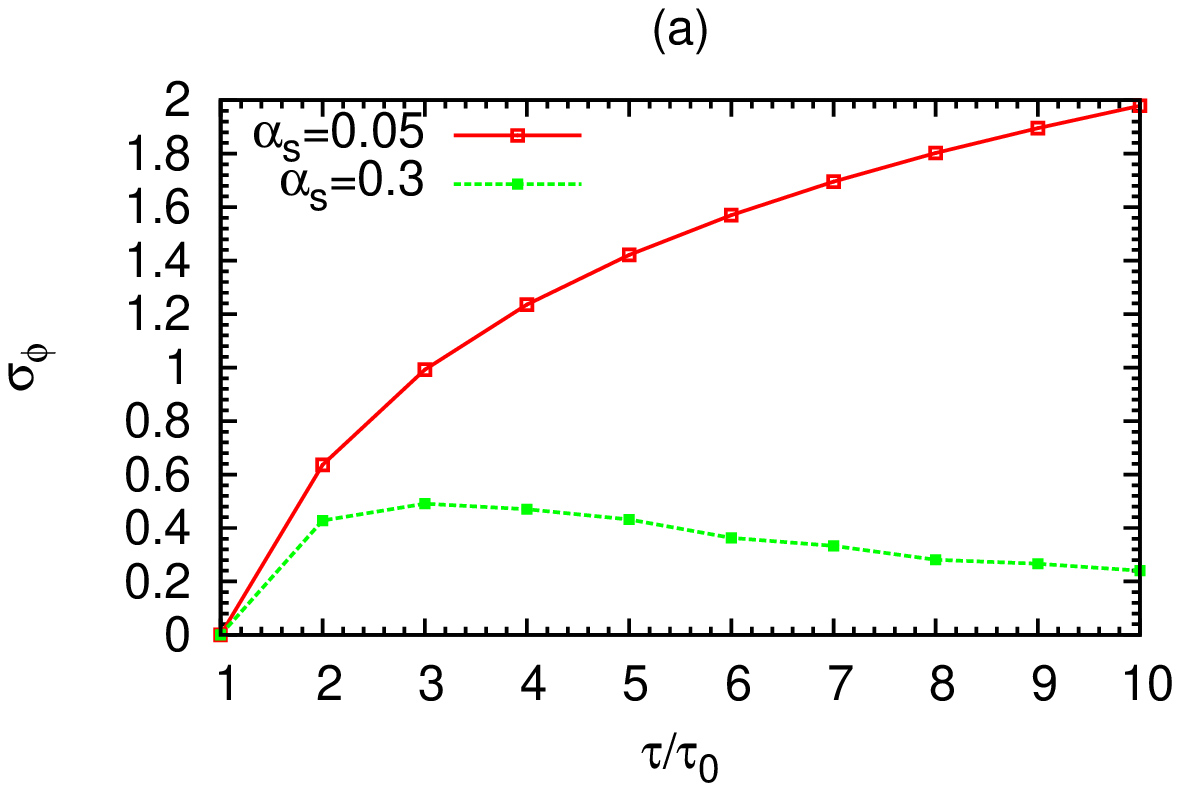}
\vfill
\epsfbox{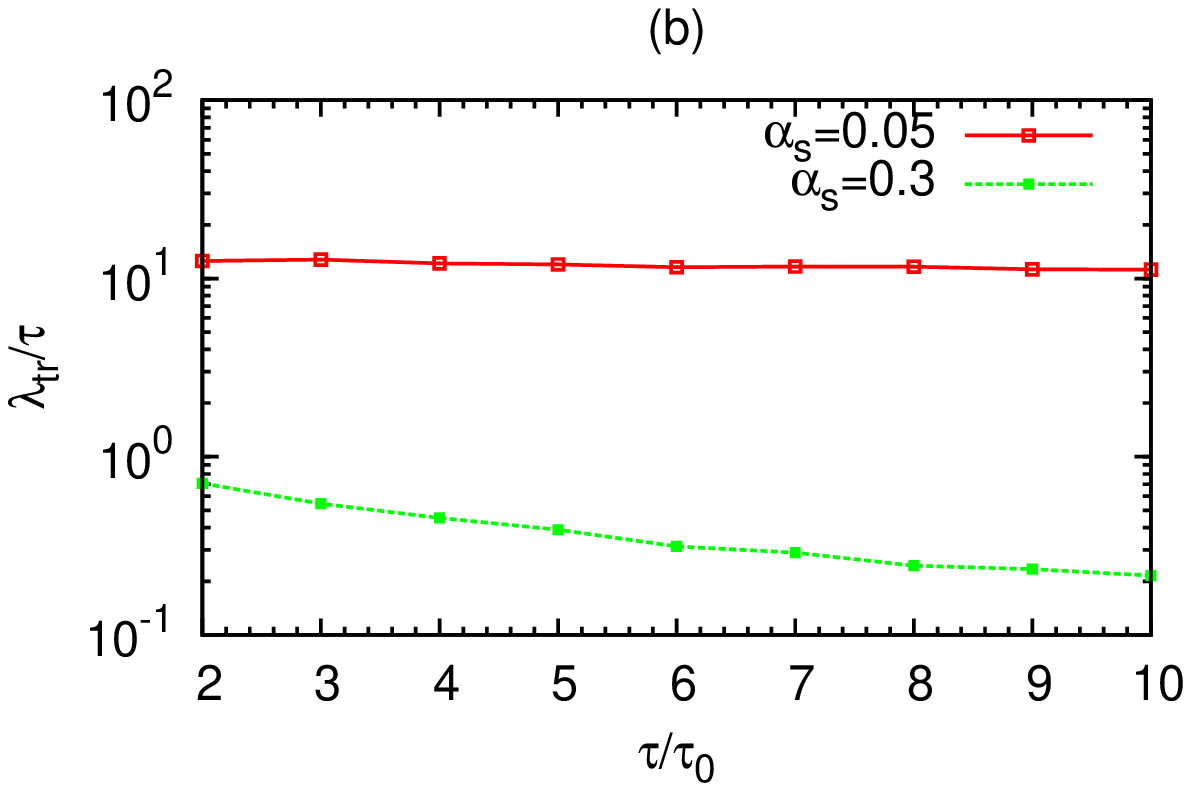}
\caption {(Color online)  (a)Variance $\sigma_\phi$ and  (b) ratio $R_{OE}$ calculated by the iterative procedure. $\tau_0=0.4$ fm/c.}
\label{sigma_R}
\end{figure}

Figure \ref{sigma_R}(a) shows the deviation from kinetic 
equilibrium, $\sigma_\phi$ from Eq. (\ref{sigmaphi}), as a function of time scaled 
with the initial time.
For $\alpha_s=0.3$ the value of $\sigma_\phi$ starts at zero (equilibrium),
increases until $3\tau_0$ and then relaxes to zero. The system first evolves
out of equilibrium and then relaxes back to equilibrium.
On the contrary, $\sigma_\phi$ increases continuously when employing a much weaker (and unphysically low) coupling $\alpha_s=0.05$.
In this case the system is always out of equilibrium. To explain the
different behaviors we define $R_{OE}$ as the ratio of the mean 
transport free path, $1/\sum R^{tr}_{Grad}$ defined by Eq. (\ref{Rtr_IR}), to
the Hubble-like expansion time scale $\tau$: 
\begin{equation}
R_{OE}=\frac{\lambda^{tr}}{\tau}=\frac{1}{\tau \cdot \sum R^{tr}_{Grad}} 
\label{R_OE}
\end{equation}
Our concept of $R_{OE}$ is similar to that introduced in \cite{HM08}, where the authors demonstrate that the ratio of expansion time to the mean free path controls the deviation from equilibrium. For a fixed $\eta_g/s_g$ the mean
transport path $\lambda^{tr}=1/\sum R^{tr}_{Grad}$ changes with time. At full equilibrium $\lambda^{tr}\sim 1/T \sim \tau^{1/3}$ and thus $\lambda^{tr}/\tau \sim \tau^{-2/3}$. If $R_{OE}$ is
larger than unity, the system starts to depart from equilibrium.  If $R_{OE}$ is
smaller than unity, the system relaxes to equilibrium. $R_{OE}(\tau)$ is shown in Fig. \ref{sigma_R}(b). 
With $\alpha_s=0.05$ the system evolves 
far away from equilibrium and the evolution is dominated by free streaming. 
The ratio $R_{OE}$ is a measure of the ability of the system to relax to kinetic equilibrium. For $\alpha_s=0.05$ kinetic equilibration is not possible for the timescales shown. The regime for which the system can not come close to kinetic equilibrium is for the coupling  $\alpha_s=0.1-0.2$ corresponding to a shear viscosity to entropy density ratio $\eta/s=0.8-0.4$. 

In addition, $\sigma_\phi$ is larger than unity at $\tau > 3\tau_0$
for $\alpha_s=0.05$. The true entropy density $s_g$ should be smaller
than that estimated according to Eq. (\ref{sdens}), because the expansion
$\ln(1+\phi)\approx \phi$ is not valid any more for large $\phi$. 
The derivation of the shear viscosity in Eq. (\ref{etacompact}) becomes 
questionable as well, since the same expansion is used to obtain the entropy 
production (\ref{ds2}). 

\begin{figure}[p]
\epsfbox{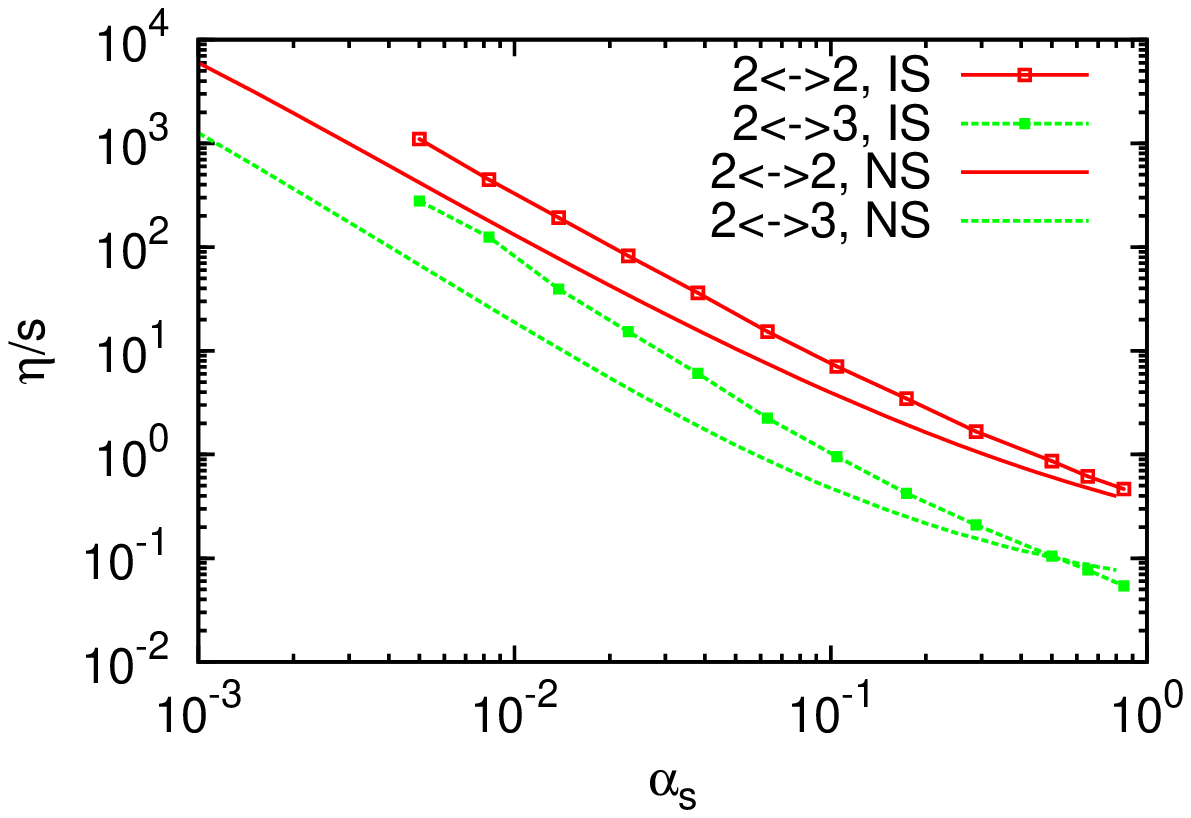}
\caption {(Color online) Ratio  $\eta/s$ (contributions due to elastic and inelastic processes) as function of the coupling constant $\alpha_s$. The result (solid line) is compared with results of  Ref. \cite{XUPRL} (dotted line)}
\label{eos_as}
\vfill
\epsfbox{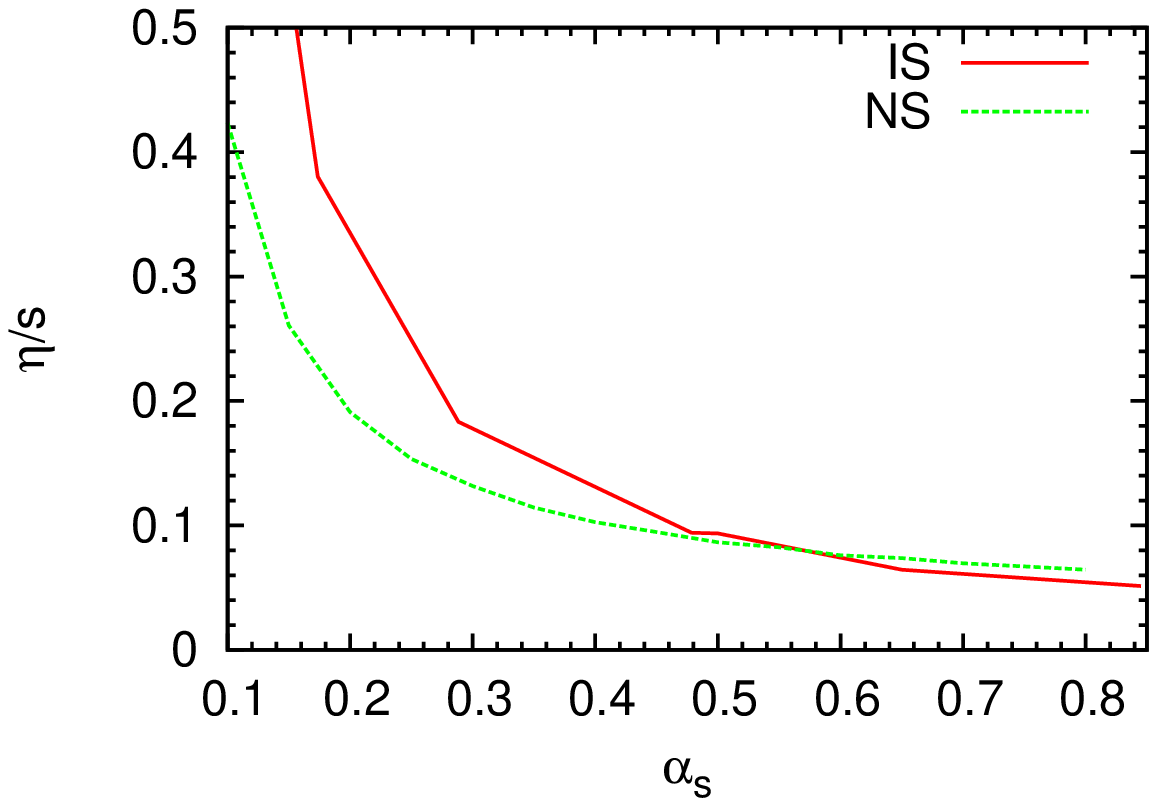}
\caption {(Color online) Ratio  $\eta/s$ (all processes) as function of the coupling constant $\alpha_s$. The result is compared with result of  Ref. \cite{XUPRL}}
\label{eos_as_all}
\end{figure}

Finally, in Figs. \ref{eos_as} and \ref{eos_as_all} we compare the results on $\eta_g/s_g$ from the second-order (IS)
kinetic theory with those presented in Ref. \cite{XUPRL} using the 
Navier-Stokes approximation. The solid (dotted) curve in Fig. \ref{eos_as} depicts the contribution of $gg\to gg$ 
($gg \leftrightarrow ggg$) to $\eta_g/s_g$ obtained in \cite{XUPRL}.
The solid (dotted) curve with symbols depicts the results from
the present calculations at $\tau=2\tau_0$, at which the system is still
near thermal equilibrium. We see that the results following from the second-order expansion are 
mostly larger than those based on the Navier-Stokes scheme,
both for $gg\to gg$ and for $gg \leftrightarrow ggg$ processes. 
At (unphysical) small $\alpha_s$ the difference between the results is given by a factor of $2-3$. 
In particular, the difference between the second-order and
the Navier-Stokes results for bremsstrahlung $gg \leftrightarrow ggg$
is bigger than that for elastic $gg\to gg$ process. At large $\alpha_s$ the $gg \leftrightarrow ggg$
processes play a dominant  role (compared with $gg\to gg$) in lowering $\eta_g/s_g$, whereas at small 
$\alpha_s$ this dominance becomes weaker \cite{AYM}. In Fig. \ref{eos_as_all} the results on $\eta/s$ implementing both elastic and inelastic processes are shown for the physical region of $\alpha_s$. Here the difference between second-order and Navier-Stokes based calculations is approximately $50\%$($\alpha_s=0.2$)-$20\%$($\alpha_s=0.3$)-$0\%$($\alpha_s=0.6$).

\section{Calculation of shear viscosity in a gluon system: transport
simulations employing BAMPS}
\label{results}
In this section, we solve the Boltzmann equation for gluons using the parton 
cascade Boltzmann Approach of MultiParton Scatterings and repeat the task 
in the previous section to calculate the shear viscosity to entropy 
density ratio $\eta_g/s_g$ in a Bjorken-type one-dimensional (0+1) expansion.
We calculate $\eta_g$ and $s_g$ according to Eqs. (\ref{etacompact})
and (\ref{sdens}) by extracting $P^{\mu\nu}$, $\bpi_g$, $n_g$ and $e_g$
from the transport simulations.

The partonic cascade BAMPS which was introduced in \cite{Xu05,Xu07} has 
been applied for a (0+1) dimensional expansion to study thermalization 
of a color glass condensate potentially produced in ultrarelativistic heavy 
ion collisions \cite{El08}. We take the same numerical setup for BAMPS as 
considered in \cite{El08}. The initial condition and interactions of 
gluons are the same as given in the previous section. In the parton 
cascade calculations, different from calculations using the viscous hydrodynamic 
equations (\ref{2nd_order_n})-(\ref{2nd_order1}), the inelastic 
$gg\leftrightarrow ggg$ processes lead to a net particle production
or absorption, i.e., $\partial_\mu N_g^\mu \ne 0$, which drives
the chemical equilibration.

We note that particle number changing processes are implemented in BAMPS, whereas the particle number was considered to be constant in previous section. Therefore we are not able to make a direct comparison between BAPMS results and those calculated by solving Israel-Stewart equations.

\begin{figure}[p]
\epsfbox{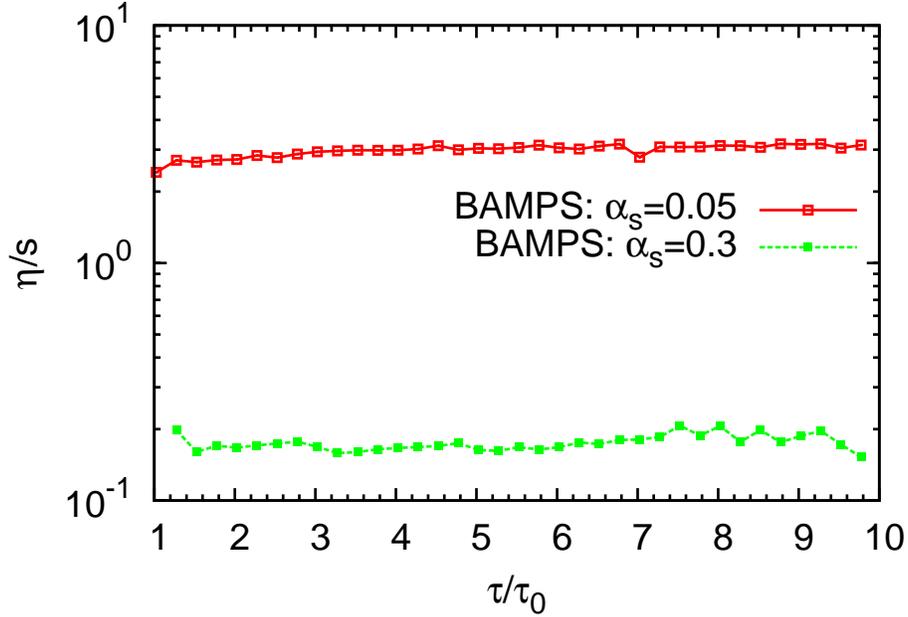}
\caption {(Color online) Ratio $\eta/s$ from the microscopic BAMPS simulation. Results are calculated using Eq. (\ref{etacompact}) for simulations with two different values for $\as$. $\tau_0=0.4~fm/c$}
\label{eos_BAMPS}
\end{figure}

\begin{figure}[p]
\epsfbox{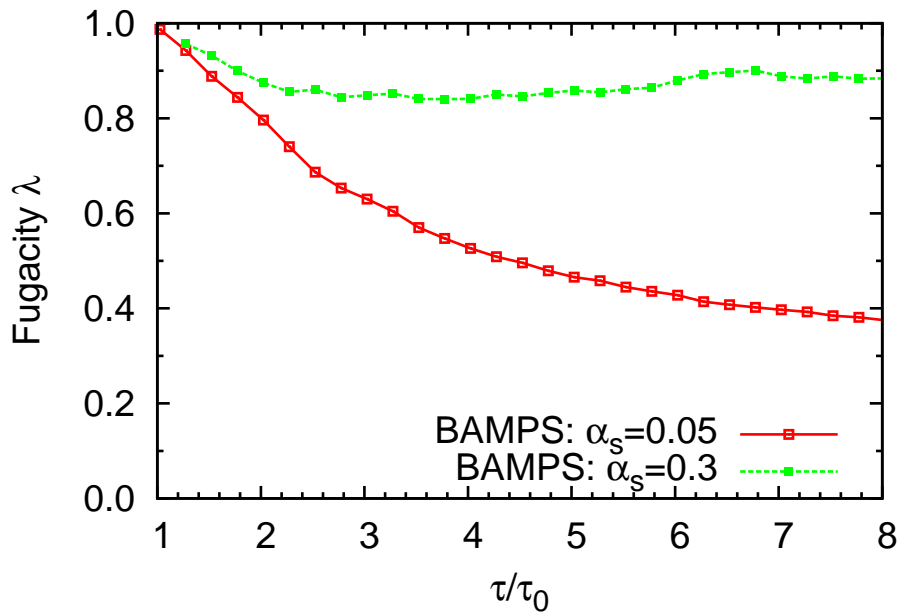}
\caption {(Color online) Fugacity $\lambda=n/n_{eq}$ from BAMPS calculation. Results are shown for simulations with different (constant) values of $\as$. $\tau_0=0.4$ fm/c.}
\label{fuga}
\end{figure}

Figure \ref{eos_BAMPS} shows $\eta_g/s_g$ extracted within the space time 
rapidity interval $\eta_s \in [-0.1:0.1]$, where 
$\eta_s=\frac{1}{2} \ln [(t+z)/(t-z)]$. When comparing these results with
those shown in the upper panel of Fig. \ref{eos_iter} we find that
they are almost the same for $\alpha_s=0.3$, whereas for $\alpha_s=0.05$
the increase of $\eta_g/s_g$ is slightly weaker in BAMPS calculations than in 
viscous hydrodynamic ones. The reason for this difference is
the different behavior of the gluon fugacity (remember that
$\eta_g/s_g \sim 1/\lambda(1-\ln \lambda)$). The gluon fugacity extracted from
BAMPS is shown in Fig. \ref{fuga}. Its value is larger than that shown in 
the lower panel of Fig. \ref{eos_iter}, because ongoing chemical equilibration
is realized in the BAMPS calculations. 

\begin{figure}[p]
\epsfbox{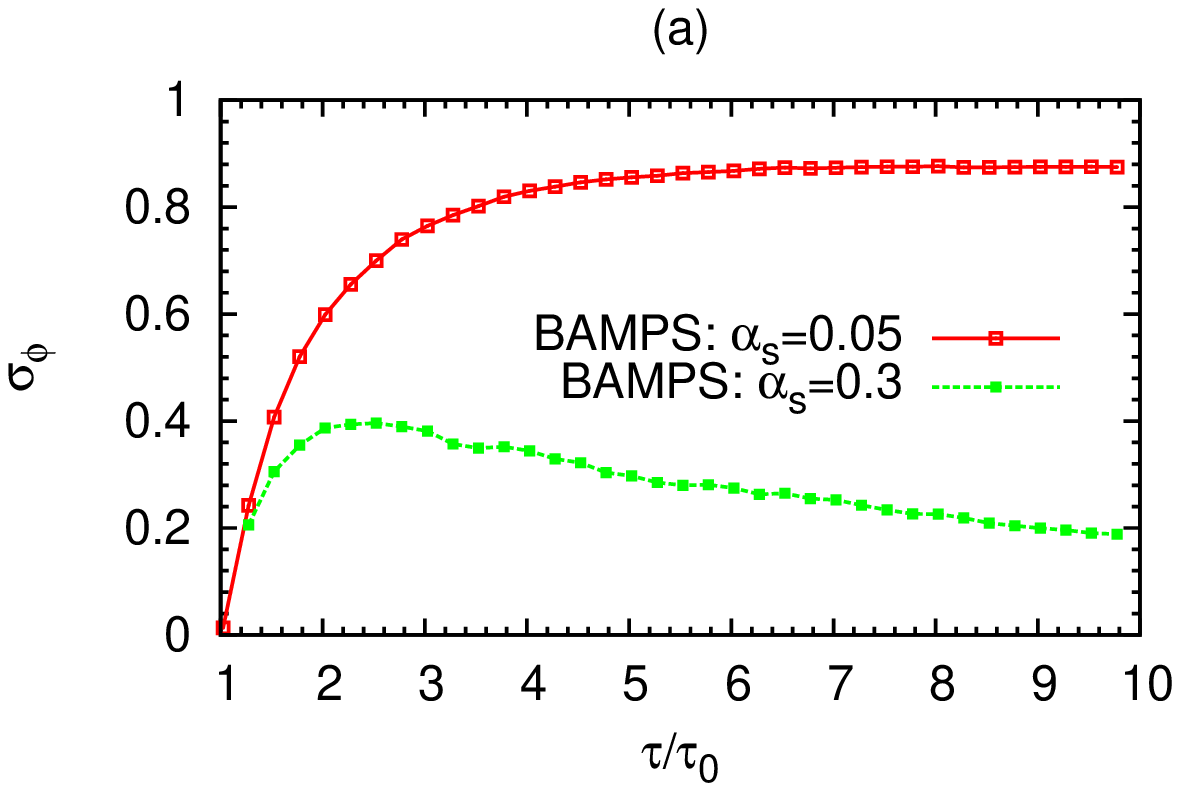}
\vfill
\epsfbox{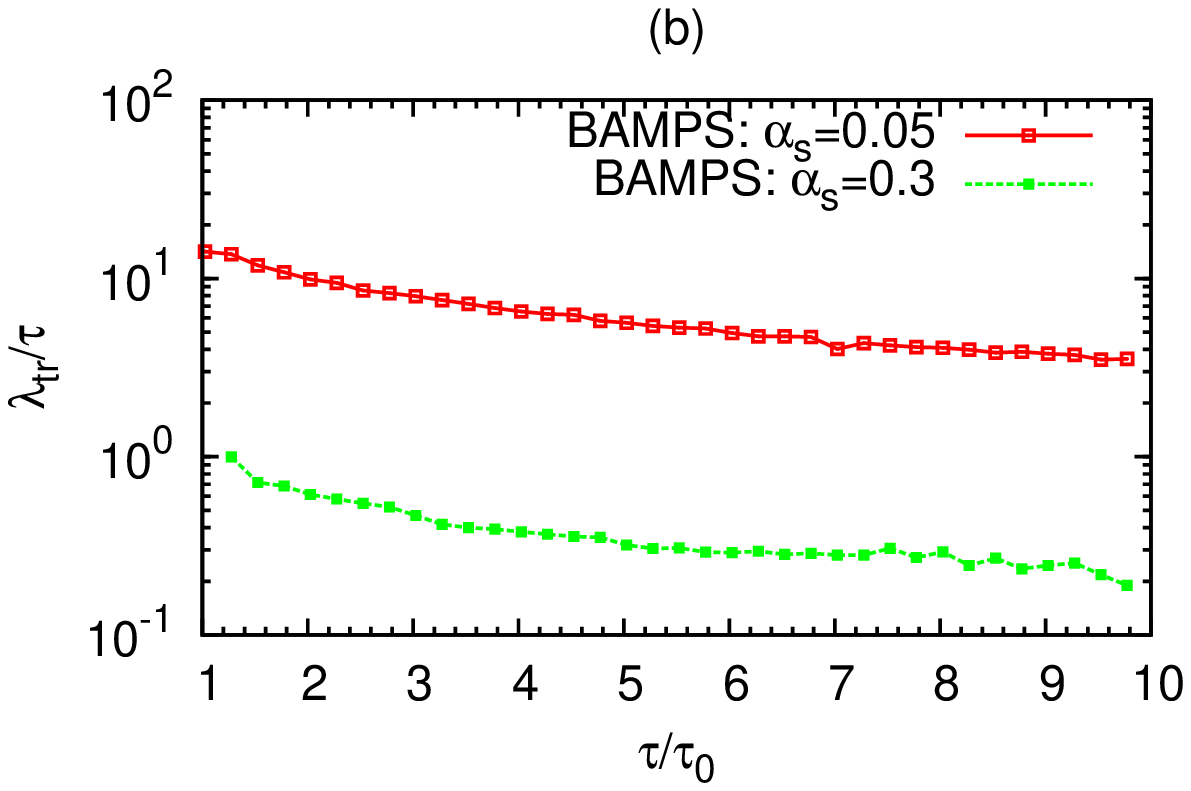}
\caption {(Color online) (a) Variance $\sigma_\phi$  and (b) ratio $R_{OE}$  calculated by BAMPS. $\tau_0=0.4$ fm/c.}
\label{sigma_R_bamps}
\end{figure}

\begin{figure}[p]
\epsfbox{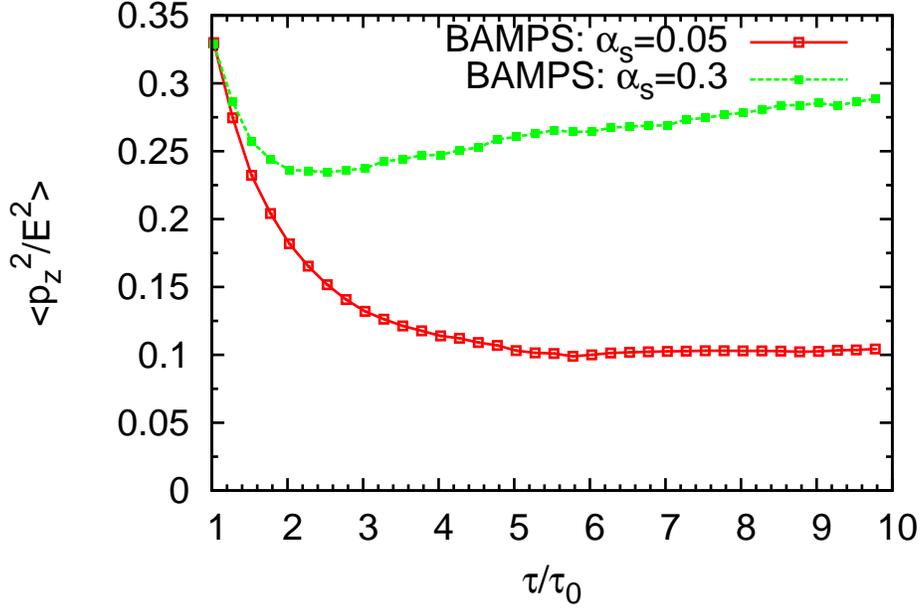}
\vfill
\caption {(Color online) Momentum isotropy $<p_z^2/E^2>$ calculated by BAMPS. $\tau_0=0.4$ fm/c.}
\label{momiso}
\end{figure}

The kinetic equilibration is demonstrated in  
Fig. \ref{sigma_R_bamps}(a) via the variance 
$\sigma_\phi$ and in Fig. \ref{momiso} via the momentum isotropy $Q(t)=<\frac{p_z^2}{E^2}>$. 
The results on $\sigma_\phi$ are similar to those in Fig. \ref{sigma_R}
and can be well understood by the out-of-equilibrium ratio $R_{OE}$ 
shown in Fig. \ref{sigma_R_bamps}(b). For $\alpha_s=0.3$ the transport mean free path is shorter than the expansion rate whereas for $\alpha_s=0.05$ the evolution is dominated by expansion. Momentum isotropization, shown in Fig. \ref{momiso}, is practically restored for $\alpha_s=0.3$ at later times, whereas for $\alpha_s=0.05$ this restoration is not possible. Here again the differences between transport and viscous hydrodynamic 
calculations stem from the different time evolution of the gluon fugacity.
To make fair comparisons, modifications in the hydrodynamic equations will be done in the near future \cite{el2} to take into account the chemical equilibration.

As pointed out already in the previous section, the parameters for which the system cannot come close to kinetic equilibrium are the couplings of $\alpha_s\approx 0.1-0.2$ corresponding to a ratio $\eta/s\sim 0.8-0.4$. For such parameters, the ratio $\lambda^{tr}/\tau$ is of the order of $1$ at late times and the system becomes highly diffusive and viscous.

\begin{figure}[p]
\epsfbox{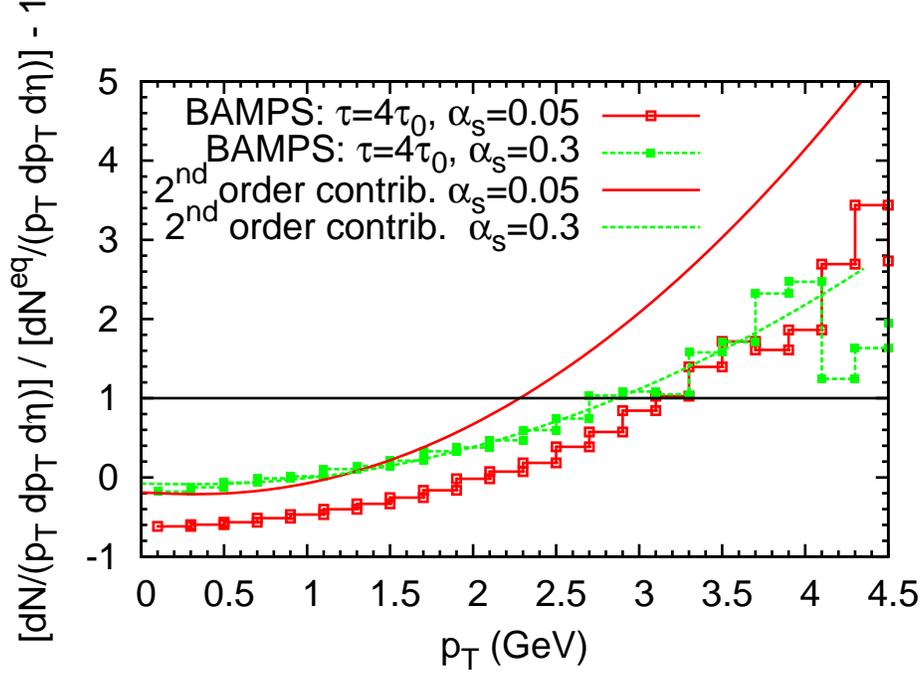}
\vfill
\caption {(Color online) Nonequilibrium part of the transverse spectrum (normalized to the equilibrium spectrum) $\frac{dN/(p_T dp_T d\eta)}{dN_{eq}/(p_T dp_T d\eta)}-1$ from BAMPS calculations (lines with points) and the second-order contribution to the transverse spectrum $\frac{\int f_{eq} \phi(\pi,T,\lambda) dp_z}{\int f_{eq} dp_z}$ (lines) as function of $p_T$ at $\tau=4\tau_0$ with $\bpi,T,\lambda$ extracted from BAMPS.}
\label{crit_pt}
\end{figure}

Finally, in Fig. \ref{crit_pt} we investigate deviations from equilibrium of the gluon distribution 
in BAMPS calculations at large momentum. Figure \ref{crit_pt} shows the non-equilibrium part of the transverse spectrum (normalized to the equilibrium spectrum) $\frac{dN/(p_T dp_T d\eta)}{dN_{eq}/(p_T dp_T d\eta)}-1$ from BAMPS calculations and the quantity $<\phi>_{p_z}(\bpi,T,\lambda) = \int f_{eq} \phi(\bpi,T,\lambda) dp_z / \int f_{eq} dp_z$ [with $\phi(x,p)=\frac{\pi^2}{8\lambda T^6}\bpi(p_z^2-\frac{1}{2}p_T^2$)], which is the analytically calculated second-order contribution to the transverse spectrum, as a function of the transverse momentum $p_T$ at $\tau=4\tau_0$. The average $<\phi>_{p_z}$ is calculated using $\bpi,T,\lambda$ extracted from the particular BAMPS calculations.  The comparison of $\frac{dN/(p_T dp_T d\eta)}{dN_{eq}/(p_T dp_T d\eta)}-1$ and $<\phi>_{p_z}$ from Fig. \ref{crit_pt} shows that for $\as=0.05$ the distribution function in BAMPS contains contributions higher order in $p_T$ and $\bpi$ and thus the second-order \textit{ansatz} (\ref{phixp_redu}) is not sufficient to describe the evolution in BAMPS. In contrast, for $\as=0.3$ the distribution function is reasonably good approximated by second-order kinetic theory over the shown momentum range. Thus we argue that the dependence of $\phi$ on $\bpi/(\lambda T^4)$ is stronger than given by \textit{ansatz} (\ref{phixp_redu}), since $\bpi/(\lambda T^4)\sim\sigma_\phi$ quantifies the strength of dissipative effects, which are stronger at $\alpha_s=0.05$. Inclusion of additional terms in Eq. (\ref{phixp_redu}) would lead to a modification of the evolution equation for $\bpi$, which follows from the conservation law for the energy momentum tensor: $0=\partial_\mu T^{\mu\nu}\sim \partial_\mu\int p^\mu p^\nu f_{eq}(1+\phi)$. 
If employing $\as=0.05$($\eta/s\approx 3$) for large $p_T > 2.3$ GeV the variance $<\phi>_{p_z}$ becomes larger than 1. For $\alpha_s=0.3$ this happens at $p_T>2.75$ GeV. For transverse momenta larger than these critical values the expansion $ln(1+\phi)\approx \phi$ done to obtain Eq.(\ref{ds2}) is invalidated. Thus in the calculation of the entropy density (and entropy production) the $\ln(1+\phi)$ term should be approximated by $\ln(1+\phi)\approx \phi-\frac{\phi^2}{2}\Theta({p_T-p_{T crit}})$, i.e. higher order terms should be taken into account in the integration over the momentum for $p_T>p_{T crit}$. However, with $\eta/s=0.18$ this correction is less than $0.5\%$, which is due to the smallness of $\sigma_\phi$. With $\eta/s=3$ the correction is $\sim 6\%$. Thus for physical values of $\eta/s\sim 0.2$ second-order hydrodynamics is valid, even though \textit{formally} breaking down at large $p_T$. In the unphysical regime $\eta/s\sim 3$ higher order corrections are not negligible. This deserves future investigation \cite{el2}.

\section{Conclusions}
\label{conclusions}
We have derived the shear viscosity coefficient $\eta$ from kinetic theory for massless particle system undergoing a one-dimensional expansion with Bjorken boost-invariance.  The derivation makes use of Grad's moment method \cite{Grad49,AM107} and is based on an expansion of the distribution function up to second order in momentum. The final expression obtained in the present work is similar to the one based on the Navier-Stokes theory \cite{XUPRL}, but the transport rate has to be calculated in a different way.  How close the result obtained using Grad's method approximates the true value determined using the Kubo-Green formula \cite{BD08} will be studied and reported in a forthcoming publication. The values needed to calculate the shear viscosity [Eq. (\ref{etacompact})] are shear tensor $\pimn$, the particle and energy densities $e$ and $n$, the fugacity $\lambda$ and finally the second moments $P_{\mu\nu}$ of the collision term from the underlying kinetic process. They can be  calculated  either using  transport setup solving the kinetic theory or from dissipative hydrodynamic (Israel-Stewart) equations (\ref{2nd_order_n})-(\ref{2nd_order1}). However, the IS equations themselves need the value of shear viscosity as a parameter. Thus we introduce a new iterative method that allows us to solve Israel-Stewart equations and calculate $\eta/s$ as a function of time and coupling constant $\alpha_s$. The results on $\eta/s$ calculated in the partonic cascade BAMPS and from IS theory are in a good agreement for physical coupling $\alpha_s=0.3$. In this regime we obtain $\eta/s=0.18$. As a further demonstration even for unphysical small coupling $\alpha_s=0.05$ the difference between BAMPS and second-order hydrodynamic calculations of $\eta/s$ is small. We obtain $\eta/s\approx 3$ in this regime. At such small coupling $\eta/s$ increases slightly in BAMPS and somewhat stronger in hydrodynamic calculations. This increase is due to the intrinsic  fugacity, which evolves differently in both calculations.

Using IS theory, we calculate $\eta/s$ ratio for a system close to equilibrium as a function of $\alpha_s$. For physical coupling $\alpha_s\approx 0.3$ the second-order result is approximately $20\%$ higher than in calculations based on first order Navier-Stokes theory \cite{XUPRL}. For $\alpha_s=0.6$ $\eta/s=0.08$ within the Israel-Stewart and Navier-Stokes prescription.
 
Deviations of hydrodynamic evolution from equilibrium are quantified in the present work introducing the variance $\sigma_\phi$ of the nonequilibrium part of the distribution function. We demonstrate that its value is smaller than unity and later decreases with time at a physical coupling $\alpha_s=0.3$ and thus our expression of $\eta$ is valid in this case. Here again hydrodynamic and BAMPS results are in good agreement. For small coupling $\alpha_s=0.05$ hydrodynamics does not relax back to equilibrium and Grad's method becomes invalid. In BAMPS in this regime the deviation of $\sigma_\phi$ from equilibrium is smaller, which is an effect of the ongoing chemical equilibration. The ability of the system to restore kinetic equlibrium is quantified by the ratio of the mean transport free path to the expansion time. We conclude that the second-order dissipative hydrodynamics is applicable in the regime $\eta/s\lesssim 0.2$ which corresponds to values of  $\alpha_s\gtrsim 0.3$. At high momenta $p_T>3$ GeV it fromally breaks down, however for $\eta/s\sim 0.2-0.4$ it is applicable even for differential observables. For really high $\eta/s\sim 3$  the applicability of hydrodynamics certainly breaks down. For the intermediate regime $0.3<\eta/s<0.8$ a more detailed analysis in the comparison of microscopic transport description to dissipative second- (or even higher) order hydrodynamics is required. 

To make consistent comparisons between the kinetic transport model BAMPS and IS solutions we have to modify the hydrodynamic equation to take into account particle production and absorption. These calculations will be reported in a forthcoming publication.

\begin{acknowledgements}
The authors thank D.~H.~Rischke and B.~Betz for helpful discussions.

A.~E. gratefully acknowledges a fellowship by the Helmholtz foundation. This work is
supported by BMBF and by the Helmholtz International Center
for FAIR within the framework of the LOEWE program (LandesOffensive zur
Entwicklung Wissenschaftlich-\"okonomischer Exzellenz) launched
by the State of Hesse. A.~M. acknowledges the support by the National Research
Foundation under Grant No. (NRF GUN   61698).

The BAMPS simulations were performed at the Center for Scientific 
Computing of Goethe University.
\end{acknowledgements}

\end{document}